\documentclass[journal=nalefd,manuscript=letter, layout=twocolumn]{achemso}

\let\oldmaketitle\maketitle
\let\maketitle\relax
\usepackage{chemformula} 
\usepackage[T1]{fontenc} 
\usepackage{graphicx}
\usepackage{dcolumn}
\usepackage{bm}
\usepackage{xcolor}

\title{Theory of photoluminescence spectral line shapes of semiconductor nanocrystals}
\author{Kailai Lin}
\email{tommy_lin@berkeley.edu}
\affiliation{Department of Chemistry, University of California, Berkeley, California
94720, USA}
\author{Dipti Jasrasaria}
\affiliation{Department of Chemistry, University of California, Berkeley, California
94720, USA}
\altaffiliation{Current address: Department of Chemistry, Columbia University, New York, New York 10027, United States}
\author{Jason J. Yoo}
\affiliation{Department of Chemistry, Massachusetts Institute of Technology, Cambridge, Massachusetts
02143, USA}
\author{Moungi Bawendi}
\affiliation{Department of Chemistry, Massachusetts Institute of Technology, Cambridge, Massachusetts
02143, USA}
\author{Hendrik Utzat}
\email{hutzat@berkeley.edu}
\affiliation{Department of Chemistry, University of California, Berkeley, California
94720, USA}
\author{Eran Rabani}
\email{eran.rabani@berkeley.edu}
\affiliation{Department of Chemistry, University of California, Berkeley, California
94720, USA}
\alsoaffiliation{Materials Sciences Division, Lawrence Berkeley National Laboratory, Berkeley, California 94720, USA}
\alsoaffiliation{The Raymond and Beverly Sackler Center of Computational Molecular and Materials Science, Tel Aviv University, Tel Aviv 69978, Israel}

\begin{document}

\twocolumn[
\begin{@twocolumnfalse}
\oldmaketitle
\begin{abstract}
Single-molecule photoluminescence (PL) spectroscopy of semiconductor nanocrystals (NCs) reveals the nature of exciton-phonon interactions in NCs. Understanding the narrow line shapes at low temperatures and the significant broadening as temperature increases remains an open problem. Here, we develop an atomistic model to describe the PL spectrum of NCs, accounting for excitonic effects, phonon dispersion relations, and exciton-phonon couplings. We use single-molecule PL measurements on CdSe/CdS core-shell NCs from $T=4$ to $T=290$~K to validate our model and find that the slightly-asymmetric main peak at low temperatures is comprised of a narrow zero-phonon line (ZPL) and several acoustic phonon sidebands. Furthermore, we identify the distinct CdSe optical modes that give rise to the optical phonon sidebands. As the temperature increases, the spectral width shows a stronger dependence on temperature, which we demonstrated to be correlated with frequency shifts and mode-mixing, reflected as higher-order exciton-phonon couplings (Duschinsky rotations). We also model the PL dependence on core size and shell thickness and provide strategies for the design of NCs with narrow linewidths at elevated temperatures. 
\end{abstract}
\end{@twocolumnfalse}
]

The optical properties of colloidal semiconductor nanocrystals (NCs) have been extensively studied over the last several decades,\cite{Bawendi1990,Brus1991,Alivisatos1996,Efros2000,Gomez2006,Klimov2007,Efros2021} leading to the development of novel optoelectronic devices.\cite{Somers2007CdSeSensors,Garcia-Santamaria2009,Wood2010,Ledentsov2011,Bronstein2015,Owen2017ChemicalDots} Many studies have aimed to understand the photoluminescence (PL) line shapes and, in particular, the dominant channels and couplings governing the homogeneous contributions to the linewidth.\cite{Bawendi1990a,Empedocles1996,Heitz1999,Gomez2006,Salvador2006,Morello2007,Sagar2008a,Sagar2008,Lin2015} In particular, low-temperature single-NC PL measurements have been commonly used to delineate the homogeneous contributions, indicating complex structures of sidebands resulting from the coupling of excitons to acoustic and optical phonons.\cite{Besombes2001,Htoon2004,Coolen2008,Fernee2008,Fernee2013,Cui2013,Cui2016} On the other hand, the exciton-phonon coupling parameters obtained in these low-temperature studies were not able to reconcile the broad, application-relevant, room-temperature emission linewidths, resulting in a lack of consensus on the strength and nature of exciton-phonon interactions in NCs. 

A thorough understanding of the structure of the NC PL spectrum and its temperature dependence calls for an atomistic, parameter-free theoretical approach that calculates the PL line shapes of experimentally relevant NC sizes. For smaller clusters, one can use first-principle methods, but these are limited by computational complexity.\cite{Palato2020} Here, we adopt such an approach that accurately describes the exciton fine structure, phonon modes, and exciton-phonon couplings of NCs.\cite{Jasrasaria2021,Jasrasaria2022a} To test and validate this approach, we compared the predictions from our theory with new, single-NC PL measurements for CdSe/CdS core-shell NCs across a wide range of temperatures, from $4$ to $290$~K, overcoming common experimental challenges, such as photo-charging, bleaching, and thermal drift.\cite{Empedocles1999} Our model yields results that match very well with the experimental measurements, reconciling the narrow, low-temperature linewidth and weak phonon sidebands with the broad, room-temperature line shape. In addition, we identify the specific phonon modes that lead to the observed phonon sidebands at low temperatures and discuss the contributions of acoustic and surface modes to the ``zero-phonon" line across different NC sizes. Furthermore, we discuss the role of dephasing on the spectral line shape and the dependence on temperature. 

We start with a model Hamiltonian that describes the ground state, a manifold of excitonic states, a bath of phonons, and the exciton-phonon couplings (expanded to lowest order in the phonon mode coordinates) for an NC,\cite{Jasrasaria2021} weakly perturbed by an electromagnetic field:
\begin{align}
H & =E_{g}\left|\psi_{g}\right\rangle \left\langle \psi_{g}\right|+\sum_{n}E_{n}\left|\psi_{n}\right\rangle \left\langle \psi_{n}\right|\nonumber \\
 & \quad+\sum_{\alpha}\hbar\omega_{\alpha}b_{\alpha}^{\dagger}b_{\alpha}+\sum_{\alpha n m}V_{n,m}^{\alpha}\left|\psi_{n}\right\rangle \left\langle \psi_{m}\right|q_{\alpha}\nonumber\\
 & \quad+\varepsilon_0 \sum_{n}\mu_{gn} \cos(\omega t)\left|\psi_{g}\right\rangle \left\langle \psi_{n}\right|+ h.c.\,. \label{eq:hamiltonian}
\end{align}
Here, $E_{g}$, $\left|\psi_{g}\right\rangle $, $E_{n}$, and $\left|\psi_{n}\right\rangle $ are the energies and wavefunctions for the ground and $n$-th excitonic states, respectively, which are obtained using the atomistic semiempirical pseudopotential method~\cite{Wang1995,Wang1996,Rabani1999} combined with the Bethe-Salpeter Equation (BSE).\cite{Rohlfing2000,Eshet2013} We used a Stillinger-Weber force field~\cite{Zhou2013} to describe the equilibrium geometry of the NC and to obtain the normal modes, which have frequencies $\omega_\alpha$ and normal mode displacements $q_{\alpha}$. The exciton-phonon couplings, $V_{n,m}^{\alpha}$, between excitonic states $\left|\psi_{m}\right\rangle $ and $\left|\psi_{n}\right\rangle $ through phonon mode $\alpha$ were also calculated from the semiempirical pseudopotential model combined with the BSE.\cite{Jasrasaria2021} In practice, we only include the diagonal coupling elements, $V_{n,n}^\alpha$, since they dominate the spectral line shape. Finally, the external electromagnetic field with strength $\varepsilon_0$ and frequency $\omega$ couples the ground state and the $n$-th excitonic state through the transition dipole $\mu_{gn}$. More details can be found in the SI and in Ref.~\citenum{Jasrasaria2021,Jasrasaria2022}.

The emission spectrum is given by the Fourier transform of the transition dipole auto-correlation function, obtained within linear response theory:\cite{mukamel1995principles} 
\footnotesize
\begin{align}
I\left(\omega\right) 
 & =\frac{1}{Q_{n}}\sum_{n}e^{-\beta E_{n}}\int_{-\infty}^{\infty}dte^{-i\omega t} \left\langle \hat{\mu}_{H}\left(t\right)\hat{\mu}_{H}\left(0\right)\right\rangle \nonumber \\
 & =\frac{1}{Q_{n}}\sum_{n}e^{-\beta E_{n}}\left|\mu_{gn}\right|^{2}\int_{-\infty}^{\infty}dte^{i\left(\omega-\omega_{ng}\right)t}\left\langle F_{n}\left(t\right)\right\rangle\,, \label{eq:many_excitons}
\end{align}
\normalsize
where we assumed a thermal Boltzmann average over initial excitonic states. In the above equation, $Q_{n}=\sum_{n} e^{-\beta E_n}$ is the partition function of excitons and $\omega_{ng}=\left( E_n-E_g\right)/\hbar$. The dephasing function, $\left\langle F_{n}\left(t\right)\right\rangle $, can be calculated using a cumulant expansion to second order,\cite{Skinner1986} and is given by the following exact expression for the model Hamiltonian (c.f., Eq.~\eqref{eq:hamiltonian}):
\footnotesize
\begin{equation}
\left\langle F_n\left(t\right)\right\rangle 
 =\exp\Bigg\{\frac{1}{2\hbar}\sum_{\alpha}\frac{\left(V_{n,n}^{\alpha}\right)^{2}}{\left(\omega_{\alpha}\right)^{3}} \left[C_\alpha^{\Re} (t) + i C_\alpha^{\Im} (t)\right]\Bigg \}\,,
 \label{eq:dephasing_function}
\end{equation}
\normalsize
where $C_\alpha^{\Re} (t) = \coth(\beta \hbar \omega_{\alpha}/2) (\cos \omega_{\alpha} t-1)$ and $ C_\alpha^{\Im}(t) = \sin \omega_{\alpha}t-\omega_{\alpha}t$. 

\begin{figure}[!ht]
\includegraphics[width=8cm]{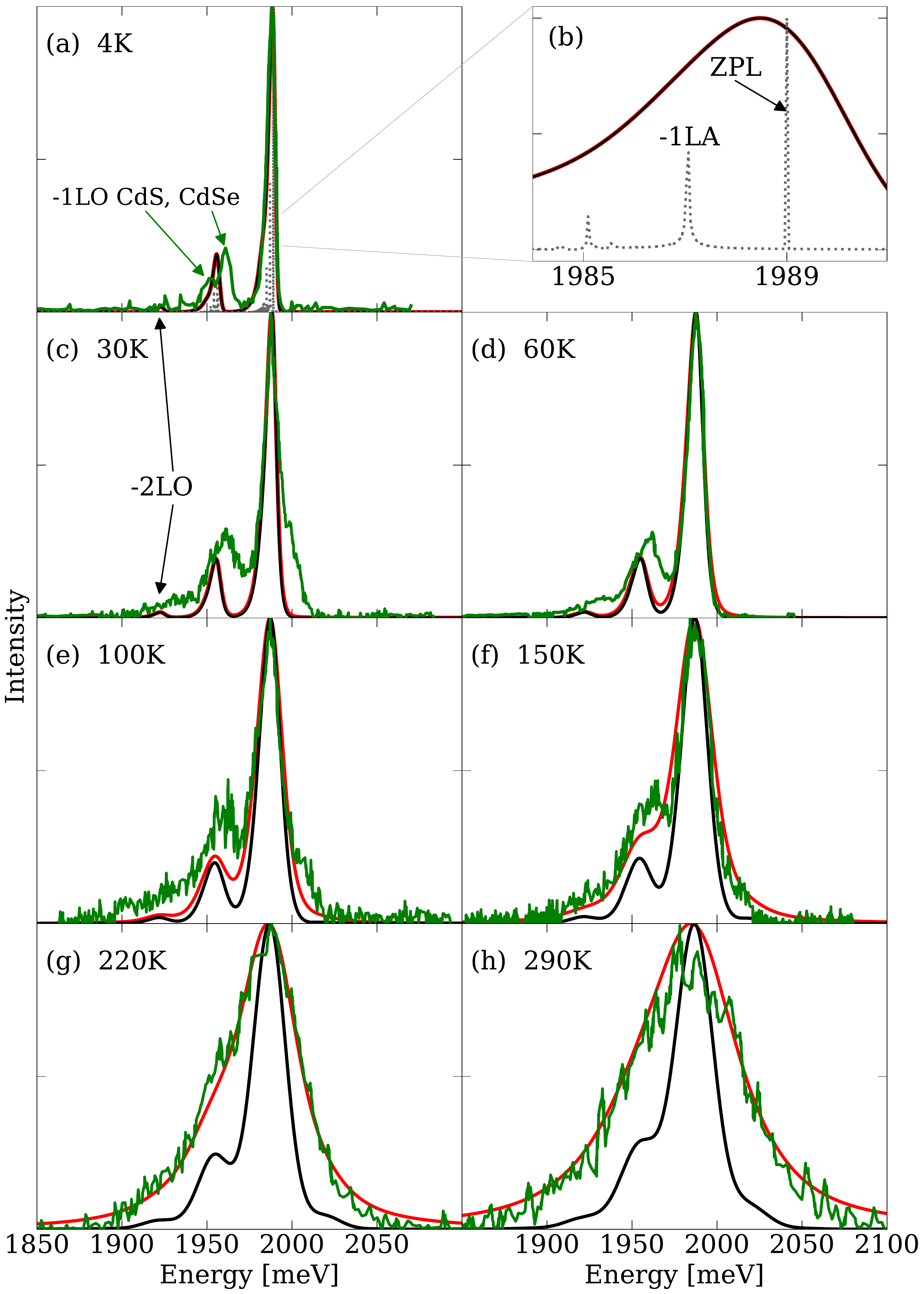}
\caption{Single-molecule photoluminescence spectra for a $3$nm diameter CdSe / $3$ML CdS core-shell NC at temperatures ranging from $4$ to $290$~K. The calculated results from the model Hamiltonian (black curves) and from the empirical inclusion of second-order expansion of exciton-phonon couplings (red curves) are compared to the experimental measurements (green curves). (a) The theoretical model results before (dotted line) and after broadening (black solid line) at $4$K are shown. Optical phonon sidebands corresponding to CdSe and CdS are distinguished. (b) A zoomed-in view around the zero-phonon line (ZPL) is shown, identifying several acoustic phonon sidebands. (c) - (h) The calculated and measured PL spectra at $30$, $60$, $100$, $150$, $220$, and $290$K are shown.}
\label{fig:Single-NC-photoluminescence-spec}
\end{figure}

In Fig.~\ref{fig:Single-NC-photoluminescence-spec}, we plot the PL spectra of a wurtzite $3$nm CdSe core / $3$ monolayer (ML) CdS shell NC for different temperatures and compare the single NC measurements (green curves) to predictions from our model (solid black curves). In comparing the experimental results with our theoretical predictions, we shifted the onset of emission due to small differences in the optical gaps ($\sim 130$meV) resulting from an additional 1-2 ML of ZnS coating in the experiments. Additionally, to compare our simulated results with experiments, we assume a Gaussian broadening of $4$meV, which accounts for experimental broadening due to the spectrometer spectral resolution and spectral diffusion. The emission shift and Gaussian broadening are the only empirical parameters used to calculate the line shapes (solid black curves) from Eqs.~\eqref{eq:many_excitons} and \eqref{eq:dephasing_function}.

The experimental spectrum at low temperatures consists of a slightly asymmetric narrow center peak and distinct phonon sidebands, which is consistent with earlier studies.\cite{Nomura1992,Empedocles1999,Besombes2001,Htoon2004,Salvador2006,Lin2015} As the temperature increases, the line shape evolves into a broader structure, and at the crossover temperature $T_c \approx 200$K, turns into a featureless, broad peak. Our theoretical model shows good agreement with the single-NC PL measurements at low and intermediate temperatures ($T \le T_c$), providing a quantitatively accurate description of the relative positions and intensities of the zero-phonon line (ZPL) and the phonon sidebands, as well as their temperature dependence. As temperature increases above $T_c$, we observe small deviations between the predicted and measured spectra, which become more significant at higher temperatures, suggesting that there is an additional emission channel that is not included in our model Hamiltonian. One such source for additional broadening is higher-order terms (beyond linear) in the expansions of the exciton-phonon couplings (Duschinsky rotations). The red curves in Fig.~\ref{fig:Single-NC-photoluminescence-spec} include an empirical correction, which is discussed further below, that accounts for the contributions of higher-order couplings at temperatures above $T_c$, resulting in excellent agreement with the experimental measurements across all temperatures. The crossover temperature is expressed in our model by equating the dephasing rate contributions from linear-order and from higher-order exciton-phonon couplings. 

\begin{figure}[!ht]
\includegraphics[width=8cm]{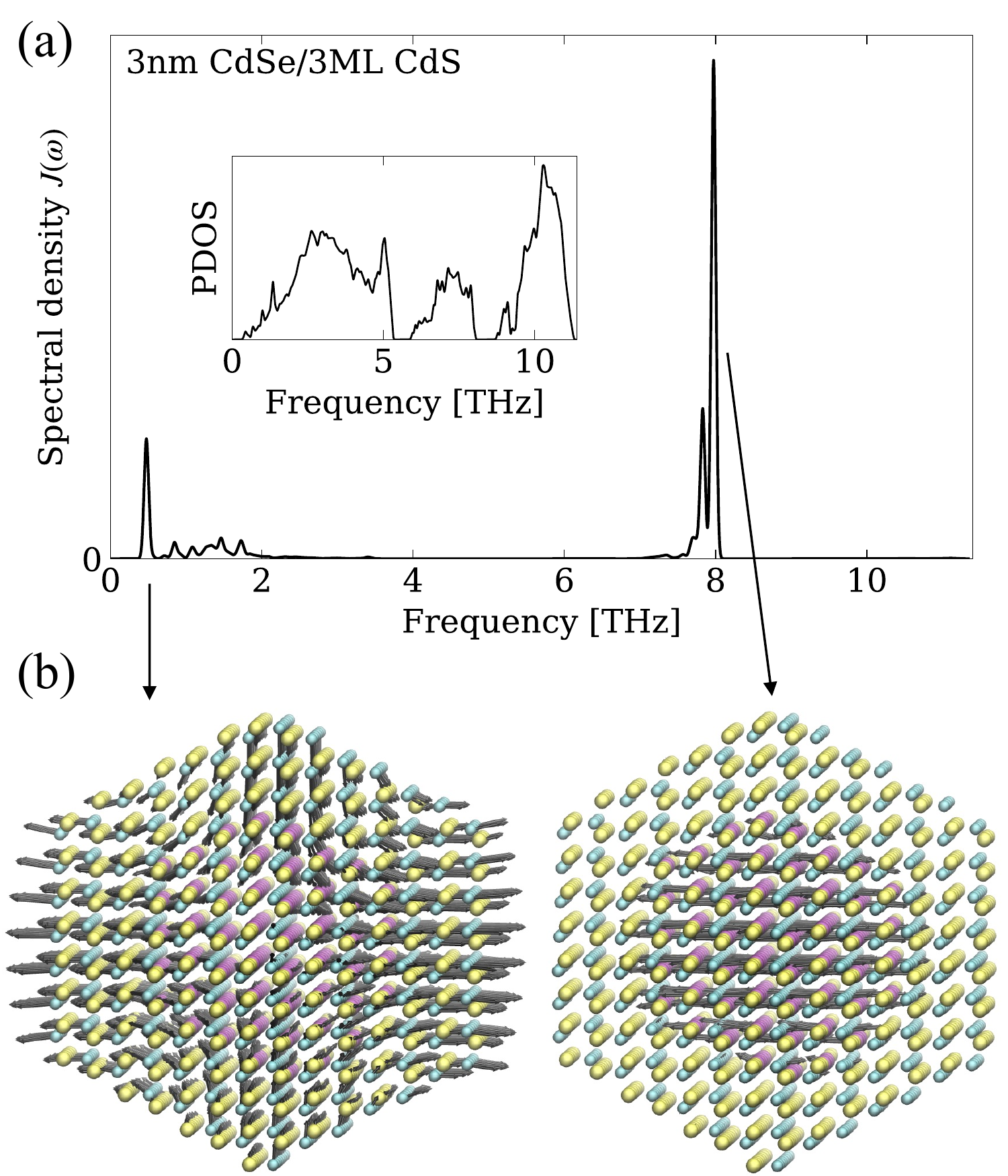}
\caption{(a) The calculated spectral density of a $3$nm core, $3$ML shell CdSe/CdS NC using our model. The inset shows the phonon density of states for this core-shell NC. (b) The motions of one acoustic phonon mode and one optical phonon with the strongest coupling to the exciton are shown. Color schemes: blue-Cd, purple-Se, yellow-S, black arrows-phonon mode motion.}
\label{fig:The-spectral-densities}
\end{figure}

In the limit that $T \rightarrow 0$K, the calculated zero-phonon line becomes infinitely narrow with transition energies corresponding to the excitonic (optical) gap,\cite{Skinner1986,Coalson1987} as shown in Fig.~\ref{fig:Single-NC-photoluminescence-spec}(b). In addition, we observe several {\em acoustic} phonon sidebands that merge with the broadened ZPL to form one main peak with a slight asymmetry that matches experiments. Our model also identifies some of the {\em optical} phonon sidebands, particularly those associated with the optical frequency of CdSe ($6.29$~THz), that remain distinct from the ZPL even at elevated temperatures ($T \le T_c$). Moreover, due to phonon emission, -2LO phonon satellite overtones are visible at energies lower than the optical transition in both the experimental and modeled spectra. The position of the CdSe LO phonon sideband is slightly shifted compared to the experiments, mainly due to inaccuracies of the classical Stillinger-Weber force field, which was parameterized for bulk semiconductors. In addition, the CdS LO phonon sideband observed experimentally at low temperatures (which otherwise merges with the CdSe LO phonon sideband) is absent in the calculated spectra. We cannot exclude emission from the charged trion state experimentally, which is known to be emissive in thick-shell CdSe/CdS, and for which we expect higher coupling to shell LO modes. 

\begin{figure*}[!ht]
\includegraphics[width=16cm]{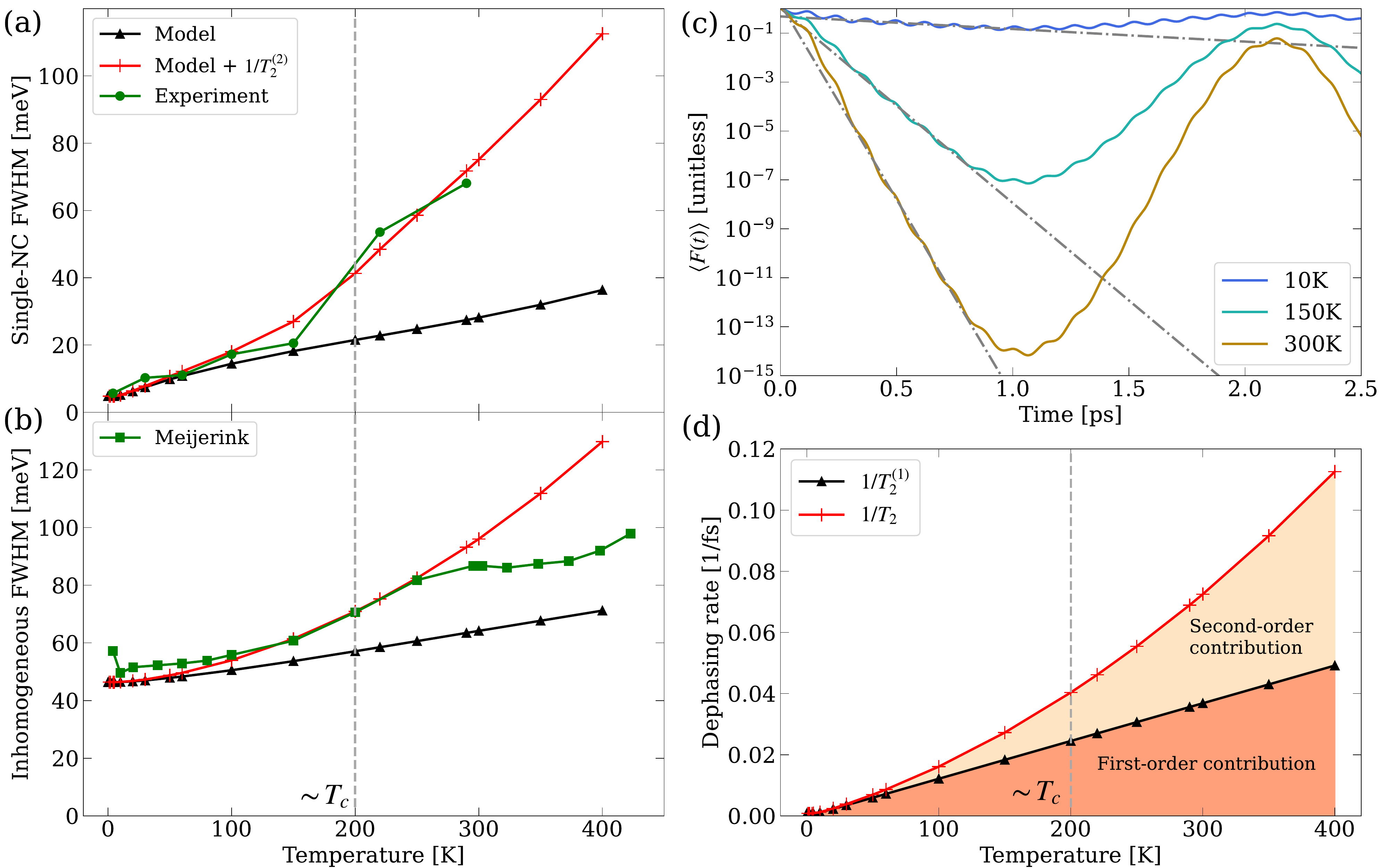}
\caption{(a) The temperature dependence of PL linewidth of single-NC PL spectra on a $3$nm/$3$ML CdSe/CdS NC. The linewidths calculated by the model (black triangles) are compared to experimental results (green circles). Dephasing contributions from second-order exciton-phonon couplings are included empirically in the red curve. (b) The temperature dependence of inhomogeneous PL linewidth in ensemble measurements on $3$nm/$2$ML CdSe/CdS NCs. Experimental results are shown in green squares and adapted from Ref.~\citenum{VanDerBok2020}. The black and red curves show the model results following the same conventions as in panel (a). (c) The dephasing function $\langle F\left( t\right)\rangle$ calculated analytically using Eq.~\eqref{eq:dephasing_function} in a semi-log plot at three representative temperatures. The first-order dephasing rates (dash-dot lines) are extracted from the exponential-decay regime of the dephasing function before recurrences happen. (d) The dephasing rate contributions from first- and second-order exciton-phonon couplings at various temperatures (see Eq.~\eqref{eq:secOrd_dephasing}). }
\label{fig:The-temperature-dependence}
\end{figure*}

To further analyze the contribution of the individual phonon modes to the spectra, we define the spectral density, which measures the strength of couplings between the exciton and the phonon modes:
\begin{equation}
J\left(\omega\right)=\sum_{\alpha}\left(\frac{V_{n,n}^{\alpha}}{\omega_{\alpha}}\right)^{2}\delta\left(\omega-\omega_{\alpha}\right)\label{eq:spectral_density}
\end{equation} 
The spectral density is plotted in Fig.~\ref{fig:The-spectral-densities}(a) for a $3$nm CdSe / $3$ML CdS NC.  Fig.~\ref{fig:The-spectral-densities}(b) shows an acoustic mode and an optical mode with the strongest coupling to the bright ground state exciton. The motions along the optical mode are primarily restricted to the CdSe core as a result of the localization of the hole to the CdSe core, while the motion along the breathing mode involves atoms in both the core and shell. Similar results were observed for other core sizes and shell thicknesses. The strong coupling to the optical phonons in the core-shell NC is consistent with earlier experimental~\cite{Scamarcio1996,Salvador2006,Lin2015,Lin2015a} and theoretical~\cite{Klein1990,Takagahara1996,Hamma2007,Kelley2011,Han2012,Han2019} investigations. 

In Fig.~\ref{fig:The-temperature-dependence}, we analyze the temperature dependence of the PL linewidth for single particle and ensemble measurements. At very low temperatures ($k_{\rm B}T<\hbar \omega_{\rm min}$), the single particle measurements (green circles) of the full width at half maximum (FWHM) are bound by the experimental resolution ($\approx 5$meV) and increase monotonically as $T$ increases, with a notable change in the slope at $T_c \approx 200$K. The FWHM for the ensemble measurements from Ref.~\citenum{VanDerBok2020}(green squares) is significantly larger than the single particle measurements due to inhomogeneous broadening, with a similar change in the temperature dependence observed at the crossover temperature, which we label as $T_c$ (see further discussion below for the physical interpretation of the crossover temperature). 

\begin{figure}[!ht]
\includegraphics[width=8cm]{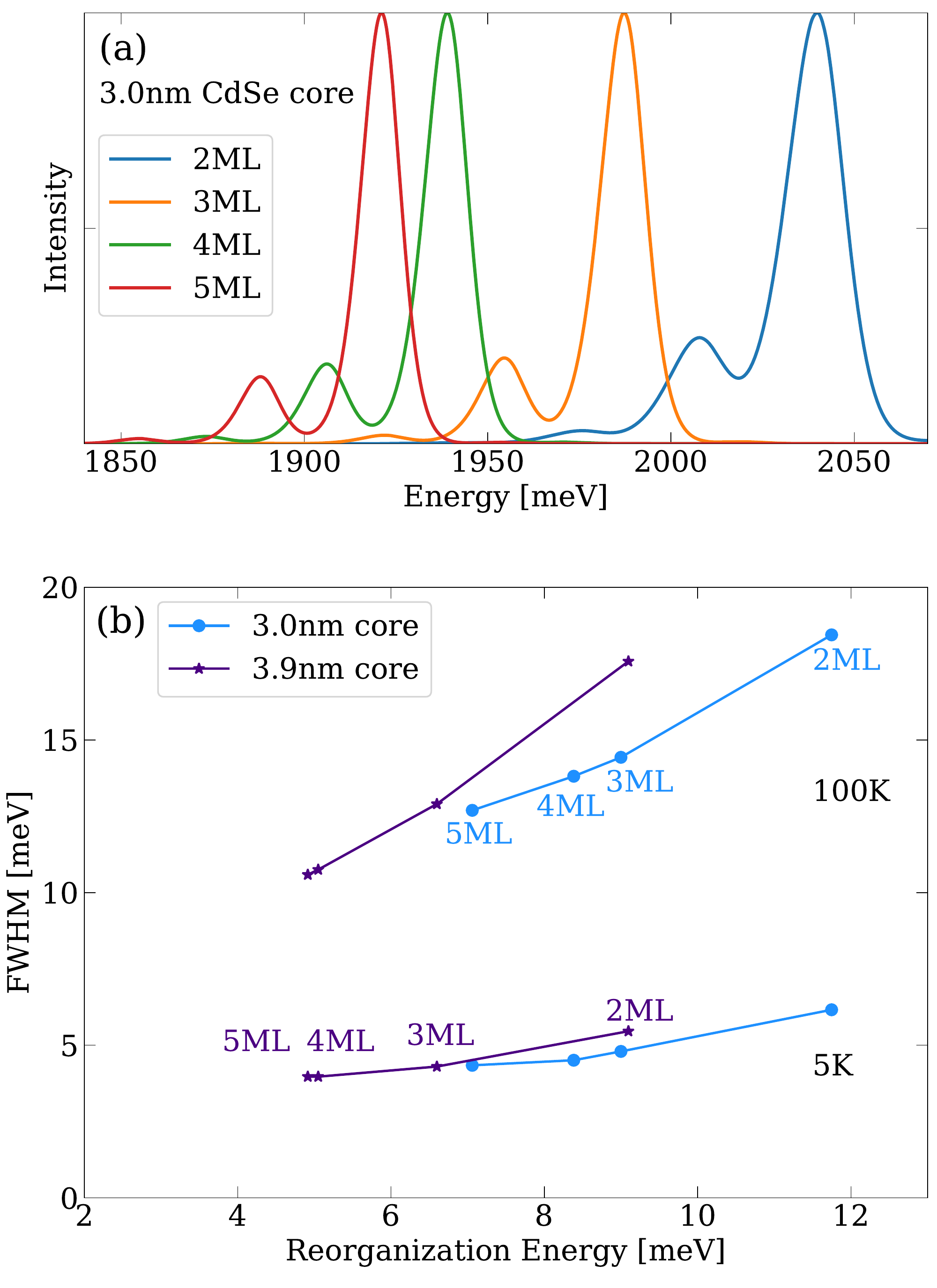}
\caption{(a) Single-NC PL spectrum progression for a CdSe/CdS core-shell NC of $3$ nm core diameter and various shell thicknesses at an intermediate temperature ($100$K). (b) The spectral linewidth vs. reorganization energy at $5$ and $100$K for NCs with $3$ nm or $3.9$ nm diameter CdSe cores and $2$ to $5$ monolayers (ML) of CdS shells. For all spectra calculations, a Gaussian broadening of $4$meV is included to account for the spectrometer spectral resolution in the experiments. }
\label{fig:Size_dependence}
\end{figure}

At low temperatures ($T \le T_c$), the results obtained from the model Hamiltonian (black curves) show excellent agreement with the experimental measurements at both the single-particle (Fig.~\ref{fig:The-temperature-dependence}(a)) and ensemble levels (Fig.~\ref{fig:The-temperature-dependence}(b)). At temperatures above $T_c$, the calculated width (black curves) shows notable deviations from the measured linewidth (green curves), as indicated above. First, we studied the anharmonicity of the atomic motion as a possible source of such deviations by evaluating the finite lifetime of phonons using classical molecular dynamics~\cite{Guzelturk2021} (see SI for more information). At $300$K, incorporating finite phonon lifetimes into the spectrum resulted in an increase of $\approx 10$meV in the single-NC FWHM, and therefore anharmonic motion is ruled out as the primary mechanism to explain the discrepancy between the experiments and theory. In addition, no temperature-activated spectral diffusion pathways have been identified,\cite{Empedocles1997,Empedocles1999,Coolen2008} suggesting that role of spectral diffusion at high temperatures is similar to that at low temperatures, where the contribution to the linewidth due to spectral diffusion is negligible. 

The discrepancy at high temperatures between the experiments and theory can also result from higher-order expansion terms in the exciton-phonon couplings. To account for these additional decay channels, we empirically correct the long-time decay rate of the dephasing function ($F(t) \propto \exp(-t/T_2)$) as follows:\cite{Skinner1986}
\begin{equation}
\frac{1}{T_2} = \frac{1}{T_{2}^{(1)}} + \frac{1}{T_{2}^{(2)}}\,,\label{eq:secOrd_dephasing}
\end{equation} 
where the dephasing rate due to the first-order coupling, 
\begin{equation}
\frac{1}{T_{2}^{(1)}} = \pi \frac{\lambda}{\hbar} \frac{k_B T}{\hbar \omega_{c}}\,,
\label{eq:perturbative_SK_firstOrd_dephasing}
\end{equation} 
is extracted from the exponential-decay of the dephasing function (c.f., Eq.~\eqref{eq:dephasing_function}), as shown in Fig.~\ref{fig:The-temperature-dependence}(c). In the above equation, the reorganization energy is defined as $\lambda = \frac{1}{2} \sum_{\alpha} \left( \frac{V_{n,n}^\alpha}{\omega_{\alpha}} \right)^2$. $\omega_{c}$ is the characteristic frequency of the bath. Assuming that the second-order coupling terms dominate the contribution from higher-order terms, the temperature dependence can be approximated by:\cite{Skinner1986}
\begin{equation}
\frac{1}{T_{2}^{(2)}} = W^2 \pi \int_{0}^{\infty} d\omega n(\omega) (n(\omega)+1)\omega^2 J^2(\omega)\,,
\label{eq:perturbative_SK_secOrd_dephasing}
\end{equation} 
where $W$ is the magnitude of second-order exciton-phonon coupling and is determined empirically. $n(\omega)$ is the Bose-Einstein distribution, and $J(\omega)$ is the spectral density defined in Eq.~\eqref{eq:spectral_density}. In Fig.~\ref{fig:The-temperature-dependence}(d), we plot the total (red curve, Eq.~\ref{eq:secOrd_dephasing}) and first-order (black curve) dephasing rates as a function of temperatures. At low temperatures, the dephasing rate is governed by the first-order term, while as the temperature increases, the contributions of the first- and higher-order terms become comparable. The temperature dependence of the total dephasing rate explains the change in slope of the FWHM as temperature increases above $T_c$, as shown by the red curves in Fig.~\ref{fig:The-temperature-dependence}(a) and (b). At temperatures around $k_B T_c = 2 / W^2 \lambda$, the first-order and second-order dephasing rates contribute equally, leading to a temperature-crossover behavior. 

Upon validating our model Hamiltonian against single-NC and ensemble PL measurements below $T_c$, we applied the model to a series of CdSe/CdS core-shell nanocrystals of various core sizes and shell thicknesses at an intermediate temperature of $100$K, at which distinct phonon sidebands are visible, as shown in Fig.~\ref{fig:Size_dependence}. The intermediate-temperature spectra are calculated with a Gaussian broadening of $4$meV that accounts for the spectrometer spectral resolution in the experiments. We observe that for the same core size and for increasing shell thickness, the energy of the optical transition decreases, as expected due to quantum confinement effects. In addition, the reorganization energy, which is a measure of the strength of exciton-phonon couplings, also decreases with increasing shell thickness and increasing core size.\cite{Jasrasaria2021, Jasrasaria2022} As a result, the widths of both the ZPL and the LO phonon sideband decrease for larger-core and thicker-shell NCs, but the effect is somewhat more pronounced for the ZPL.

Panel (b) of Fig.~\ref{fig:Size_dependence} shows the dependence of the zero-phonon linewidth on the shell thickness and the reorganization energy, respectively. The reorganization energy correlates with measured Stokes shifts,\cite{Jasrasaria2022a,Liptay2007} and thus, can also be inferred from experiments. At $100$K, the zero-phonon linewidth decreases by $\approx 6$meV as the shell thickness increases from $2$ to $5$ monolayers for both core sizes and seems to saturate beyond $\approx 5-6$ monolayers for the larger core. At $5$K, the ZPL linewidth also shows a similar linear dependence on the reorganization energy, with a less steep slope and smaller change in FWHM ($2-3$meV as the shell thickness increases). Given that the first-order exciton-phonon coupling dominates the dephasing rate $1/T_2$ below $T_c \approx 200$K, the near-linear dependence of the FWHM on the reorganization energy at $100$K and $5$K can be explained by Eq.~\eqref{eq:perturbative_SK_firstOrd_dephasing} where the first-order dephasing rate depends linearly on reorganization energy. 

In this work, we developed an approach to calculate the photoluminescence spectra of CdSe/CdS core-shell nanocrystals and compared our prediction with single-molecule PL measurements for a wide range of temperatures, overcoming challenges of blinking and rapid charging-induced spectral shifts. Our approach utilizes an atomistic force field to model the phonon modes, and the semiempirical pseudopotential model combined with the Bethe-Salpeter Equation to describe the excitonic structure and the exciton-phonon couplings to lowest order in the phonon modes in NCs.

The single-NC experiments reveal low-temperature spectra of a narrow main peak accompanied by distinct phonon sidebands and broad, featureless room-temperature spectra. Using linear response theory, we showed that the calculated results accurately reproduce our measured PL spectra below a crossover temperature, $T_c\sim 200$\,K. In the low-temperature regime, our model explains the nature of the narrow, slightly-asymmetric zero-phonon line. By analyzing the coupling of individual phonon modes to the ground state bright exciton, we identified the specific acoustic and optical phonon modes that lead to the sidebands. 

As temperature increases above $T_c$, the measured zero-phonon linewidth increases more rapidly with temperature. Anharmonic atomic motion is examined and ruled out as the source of further broadening in PL linewidth. We attributed this crossover behavior to the increasing significance of higher-order exciton-phonon coupling, which is not included in our model, and showed that including a second-order coupling term with a single parameter accounts for the behavior of the spectral linewidth above $T_c$. 

We applied our model to predict the behavior of the PL spectra for CdSe/CdS NCs of different core sizes and shell thicknesses and found that NCs with larger cores and thicker shells have smaller overall exciton-phonon coupling, as demonstrated by their reorganization energies, as well as narrower ZPL and phonon sidebands. The theoretical approach established in this work assists future designs of novel semiconductor nanocrystal materials. Increasing core size, increasing shell thickness, reducing exciton-acoustic-phonon coupling, and reducing second-order exciton-phonon coupling (Duschinsky rotations) have all been demonstrated as viable paths to achieve narrow room-temperature linewidth in relevant NC-based technologies. 

\begin{acknowledgement}
E.R. acknowledges support from the U.S. Department of Energy, Office of Science, Office of Basic Energy Sciences, Materials Sciences and Engineering Division, under Contract No. DE-AC02-05-CH11231 within the Fundamentals of Semiconductor Nanowire Program (KCPY23). Methods used in this work were provided by the Center for Computational Study of Excited State Phenomena in Energy Materials (C2SEPEM), which is funded by the U.S. Department of Energy, Office of Science, Basic Energy Sciences, Materials Sciences and Engineering Division, via contract no. DE-AC02- 05CH11231, as part of the Computational Materials Sciences Program. Computational resources were provided in part by the National Energy Research Scientific Computing Center (NERSC), a U.S. Department of Energy Office of Science User Facility operated under contract no. DE-AC02- 05CH11231. D.J. acknowledges the support of the Computational Science Graduate Fellowship from the U.S. Department of Energy under grant no. DE-SC0019323.
\end{acknowledgement}

\begin{suppinfo}
Further details are provided on nanocrystal synthesis, single-nanocrystal spectroscopy, theoretical methods for parametrizing the nanocrystal model hamiltonian, the linear response model for vibronic spectrum, and anharmonicity. See xxx [link]. 
\end{suppinfo}

\begin{tocentry}
\begin{center}
\includegraphics[width=3.25in]{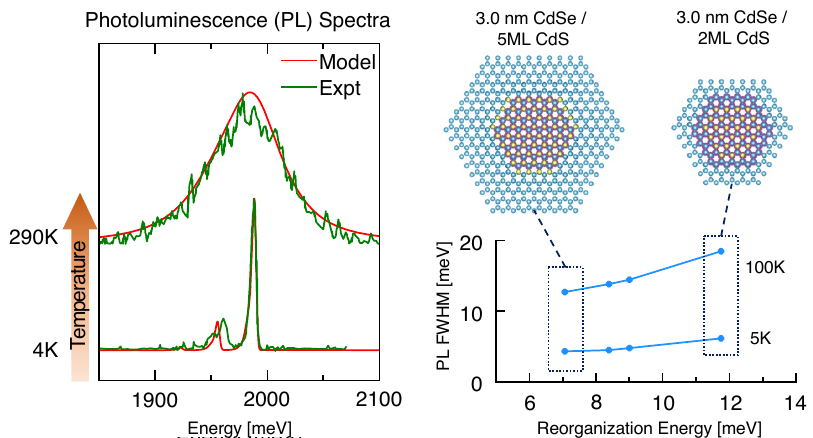}
\end{center}
\end{tocentry}


\providecommand{\latin}[1]{#1}
\makeatletter
\providecommand{\doi}
  {\begingroup\let\do\@makeother\dospecials
  \catcode`\{=1 \catcode`\}=2 \doi@aux}
\providecommand{\doi@aux}[1]{\endgroup\texttt{#1}}
\makeatother
\providecommand*\mcitethebibliography{\thebibliography}
\csname @ifundefined\endcsname{endmcitethebibliography}
  {\let\endmcitethebibliography\endthebibliography}{}

\end{document}


\tableofcontents
\section{Nanocrystal synthesis}
The synthesis of the CdSe cores and the first fast-injection shell growth was carried out by following a previously published procedure by Carbone et al.\cite{Carbone2007} For the CdS shell growths, to a 250mL round bottom flask was added 100nmol of the CdSe cores dissolved in hexane, 3mL of ODE, 3mL of oleylamine, and 3mL of oleic acid. The solution was degassed at r.t. for 2 hr. and then for 5 min at 100$^\circ C$ to remove the hexane and water. The solution was then stirred under N2 and the temperature was set to rise to 310$^\circ C$. Once the temperature reached 200$^\circ C$, a solution of Cd(OA)2 in ODE (0.08M) and a separate solution of octanethiol dissolved in ODE (0.096M) were injected at a rate of 2.5mL/hr. Finally, the CdSe/CdS NCs were overgrown with a ZnS shell. For this, 0.1M Zn(OA)2 and 0.12 M octanethiol/ODE were injected at 3 ml/hr to the CdSe/CdS NCs for the appropriate amount of time to achieve desired shell thickness. After injection, the solution is further annealed at 310$^\circ C$ for 15 min. 

\section{Single-nanocrystal spectroscopy}
Samples for single molecule spectroscopy were prepared by diluting the stock-solution of nanocrystals in a PMMA solution (3\% by weight in toluene), and spincoating (5000 rpm, 1 min) the solution on quartz slides (MTI, optical grade). The samples were cooled to liquid helium temperatures with a closed cycle liquid helium cryostat (Cryostation, Montana Instruments). To adjust the temperature, Montana Instrument\textquoteright s Agile Temperature Sample Mount (ATSM), which allows fast and reliable adjustment of the temperature without too much spatial drift. Single particle spectroscopy was performed using a home-built confocal fluorescence microscope. Single nanocrystals were excited with a 488nm CW laser (Coherent Sapphire) for the recording of the temperature-dependent emission spectra and single nanocrystal temperature-dependent lifetimes. The single particle emission was filtered spatially (telescope (8 cm focal length with a 30 $\mu m$ pinhole) and spectrally (dichroic notch (488nm, Edmund Optics) and longpass filters (500nm, Thorlabs)). Single particle spectra were recorded by diffracting the mission in a monochromator (Acton 2050, Princeton Instruments) and detection with a cooled EMCCD camera (ProEM512B, Princeton Instruments). For the PL lifetime measurements, single nanocrystals were excited non-resonantly with a picosecond diode laser (PicoQuant, LDH-C-400 - 400 nm). The emission was collected in time-tagged mode (T3 mode, Picoquant Hydraharp). Photon-detection was performed with an APD (Excelitas) with a nominal timing resolution of FWHM 200ps.

\section{Nanocrystal Configurations}
The configurations of the core-shell CdSe/CdS NCs in theoretical calculations were constructed by first cutting the desired core size from bulk wurtzite CdSe geometry, and then adding several monolayers of shell (CdS). The core-shell geometries were relaxed using the conjugate gradient minimization with the classical Stillinger-Weber force field, which was parametrized for the bulk CdSe and CdS systems.\cite{Zhou2013} Upon geometry relaxation, the outermost layer of atoms were removed and replaced with ligand pseudopotentials representing the passivation layer.\cite{Rabani1999,Jasrasaria2022a} The molecular formulae, core diameters, shell thicknesses and overall diameters are summarized below in Table~\ref{table:S1} for all the CdSe/CdS core–shell NCs calculated in this work. 

\begin{table}[h!]
\centering
\begin{tabular}{c c c c} 
 \hline
 Configuration & Core diameter (nm) & Shell thickness (ML) & Overall diameter (nm)\\ [0.5ex] 
 \hline\hline
 $\mathrm{Cd_{753}Se_{252}S_{501}}$ & 3.0 & 2 & 4.6 \\ 
 \hline
 $\mathrm{Cd_{1197}Se_{252}S_{945}}$ & 3.0 & 3 & 5.4 \\
 \hline
 $\mathrm{Cd_{1788}Se_{252}S_{1536}}$ & 3.0 & 4 & 6.2 \\
 \hline
 $\mathrm{Cd_{2547}Se_{252}S_{2295}}$ & 3.0 & 5 & 7.0 \\
 \hline
 $\mathrm{Cd_{1197}Se_{483}S_{714}}$ & 3.9 & 2 & 5.4 \\ 
 \hline
 $\mathrm{Cd_{1788}Se_{483}S_{1305}}$ & 3.9 & 3 & 6.2 \\
 \hline
 $\mathrm{Cd_{2547}Se_{483}S_{2064}}$ & 3.9 & 4 & 7.1 \\
 \hline
 $\mathrm{Cd_{3495}Se_{483}S_{3012}}$ & 3.9 & 5 & 7.9 \\
 \hline
\end{tabular}
\caption{CdSe/CdS core-shell NC geometry details.}
\label{table:S1}
\end{table}

\section{Theoretical methods for semiconductor nanocrystals}
As stated in the main text, we use a model Hamiltonian to describe the excitonic states, phonon modes, and exciton-phonon interactions within a semiconductor nanocrystal, which is also weakly perturbed by an external electric field (See Eq.(1) from the main text). To parameterize this Hamiltonian within reasonable computational cost, we resort to the atomistic semiempirical pseudopotential method~\cite{Wang1995,Wang1996,Rabani1999} combined with the Bethe-Salpeter Equation (BSE).\cite{Rohlfing2000,Eshet2013} We also use a Stillinger-Weber force field~\cite{Zhou2013} to describe the equilibrium geometry of the NC and to obtain the phonon modes. This pseudopotential-BSE approach and the set of parameters for CdSe and CdS have been shown to yield accurate descriptions of the exciton fine structure in CdSe/CdS core-shell nanocrystals systems.\cite{Rabani1999,Philbin2018,Jasrasaria2022a,Jasrasaria2021}

The atomistic semi-empirical pseudopotential approach, adapted from Ref.~\citenum{Jasrasaria2022a}, was used to calculate the exciton fine structure of the CdSe/CdS nanocrystals. A non-interacting electron picture was used. The pseudopotential was assumed to have a local form in the reciprocal space, with a pre-factor expanded up to cubic order in the trace of the local strain tensor to account for deformation potential coupling:\cite{Jasrasaria2022a,Rabani1999,Wang1999}
\begin{equation}
\left\{ -\frac{1}{2}\nabla^{2}+\sum_{\mu}\hat{v}_{\mu}\left(\mathbf{r}-\mathbf{R}_{\mu}\right)\right\} \psi_{i}=E_{i}\psi_{i}\,, \label{eq:non-interacting-electron-hamiltonian}
\end{equation}
\begin{equation}
\hat{\tilde{v}}_{\mu}\left(\mathbf{k}\right) =\left[1+a_{4}Tr\epsilon_{\mu}+a_{5}\left(Tr\epsilon_{\mu}\right)^{3}\right] \cdot\frac{a_{0}\left(k^{2}-a_{1}\right)}{a_{2}e^{a_{3}k^{2}}-1}\,, \label{eq:pp-form}
\end{equation}
where $\hat{v}_{\mu}\left(\mathbf{r}\right)$ and $\hat{v}_{\mu}\left(\mathbf{k}\right)$ are the pseudopotential around atom $\mu$ in the real and reciprocal space, respectively. $\epsilon_{\mu}$ is the local strain tensor around atom $\mu$, and its trace is calculated as the tetrahedron volume formed by the nearest neighbors. Parameters $a_0$ through $a_3$ were fitted to the band structures of bulk CdSe and CdS obtained empirically or through high-accuracy electronic structure calculations, such as DFT+GW.\cite{Cohen1966,Hybertsen1986} The local strain-dependent prefactors ($a_4$ and $a_5$) were fitted to the bulk deformation potentials of the CBM and VBM for each semiconductor material.\cite{Jasrasaria2022a,Li2006} After the single-electron Hamiltonian is constructed, the filter diagonalization technique was used to obtain the quasi-particle states (electron or hole states) near the band edge of the nanocrystal.\cite{Neuhauser2000,Toledo2002} The pseudopotential parameters used in the work for Cd, Se and S atoms are given in Table~\ref{table:S2}.\cite{Jasrasaria2022a}

\begin{table}[h!]
\centering
\begin{tabular}{c c c c c c c} 
 \hline
  & $a_0$ & $a_1$ & $a_2$ & $a_3$ & $a_4$ & $a_5$\\ [0.5ex] 
 \hline\hline
 Cd & -31.4518 & 1.3890 & -0.0502 & 1.6603 & 0.0586 & 0 \\ 
 \hline
 Se & 8.4921 & 4.3513 & 1.3600 & 0.3227 & 0.1746 & -33 \\
 \hline
 S & 7.6697 & 4.5192 & 1.3456 & 0.3035 & 0.2087 & 0 \\
 \hline
\end{tabular}
\caption{Pseudopotential parameters for Cd, Se and S. All parameters are given in atomic units}
\label{table:S2}
\end{table}

Excitonic states were expressed as a linear combination of electron-hole product wavefunctions, where the single-particle states are correlated. The coefficients ($c_{a,i}^{n}$ in the equation below) were obtained by solving the Bethe-Salpether equation within the static screening approximation.\cite{Rohlfing2000} 
\begin{equation}
\psi^{n} \left(\mathbf{r}_e, \mathbf{r}_h\right) = \sum_{a,i} c_{a,i}^{n} \phi_a \left(\mathbf{r}_e\right) \phi_i \left( \mathbf{r}_h\right) \,, \label{eq:BSE}
\end{equation}
where $\psi^{n} \left(\mathbf{r}_e, \mathbf{r}_h\right)$ is the exciton wavefunction. $\phi_a \left(\mathbf{r}_e\right)$ and $\phi_i \left( \mathbf{r}_h\right)$ are the single-particle electron and hole states obtained from the filter diagonalization technique, as discussed above. To converge the ground exciton state energy for the NC structures in this work, 60 electron and 60 hole states near the band edge were used.

The exciton-phonon coupling matrix element to first order was calculated non-adiabatically as the derivative of semi-empirical pseudopotential, following the methods and developments shown in Ref.~\citenum{Jasrasaria2021}. The exciton-atomic-coordinate coupling $V_{n,m}^{\mu k}$ has contributions from the electron and the hole channels, where the coefficients from the Bethe-Salpether equation are factored in: 
\begin{equation}
V_{n,m}^{\mu k}=\sum_{abi}c_{ai}^{n}c_{bi}^{m}v_{ab,\mu}^{\prime}-\sum_{aij}c_{ai}^{n}c_{aj}^{m}v_{ij,\mu}^{\prime} \,, \label{eq:Vkl}
\end{equation}
where $v_{rs,\mu}^{\prime}=\left\langle \phi_{r}\left|\left(\frac{\partial v_{\mu}\left(\left|\mathbf{r}-\mathbf{R}_{\mu}\right|\right)}{\partial R_{\mu k}}\right)_{\mathbf{R}_{0}}\right|\phi_{s}\right\rangle $, $c_{ai}^{n}$ and $c_{bj}^{m}$ are coefficients from the Bethe-Salpether equation, $\phi_{r}$ and $\phi_{s}$ are the single-particle states. The coupling matrix $V$ was then transformed from the atomic-coordinate $\mu k$ to the phonon mode coordinates $\alpha$ through the eigenvectors of the mass-weighted Hessian matrix, yielding the exciton-phonon coupling matrix. 
\begin{equation}
V_{n,m}^{\alpha}=\sum_{\mu k}\frac{1}{\sqrt{m_{\mu}}}E_{\mu k,\alpha}V_{n,m}^{\mu k} \,, \label{eq:Vklq}
\end{equation}
where $E_{\mu k,\alpha}$ is the $\mu k$ element of the $\alpha$ eigenvector of the mass-weighted Hessian matrix \textbf{D}, which was constructed by taking the second-order derivative of the Stillinger-Weber potential.

\section{Linear response model for photoluminescence spectrum}
Our photoluminescence linewidth model combines the information from exciton fine structure, phonon modes, and exciton-phonon couplings to calculate the emission spectrum within linear response and the dipole approximation.\cite{mukamel1995principles} The emission spectrum is calculated as the Fourier transform of the dipole auto-correlation function in the time domain. We also adopt the Condon approximation, which assumes that the dipole operator couples the ground and excited states. We first consider only the transition between the lowest excitonic state $\left|\psi_{1}\right\rangle$ and the ground state $\left|\psi_{g}\right\rangle$, and then extend our model to all emission transitions. 
\begin{equation}
\hat{\boldsymbol{\mu}}=\boldsymbol{\mu}\left(\left|\psi_{g}\right\rangle \left\langle \psi_{1}\right|+\left|\psi_{1}\right\rangle \left\langle \psi_{g}\right|\right) \,,
\end{equation}
The dipole auto-correlation function was then calculated by taking a partial trace over the electronic (system) degrees of freedom, resulting in an average over the nuclear/phonon (bath) degrees of freedom. The last term in Eq.~\eqref{eq:trace} can be rewritten in the form of a time-ordered exponential, also known as the dephasing function: \cite{Skinner1986,Egorov1997} 
\begin{align}
C_{\mu\mu}\left(t\right) & =\left\langle \hat{\boldsymbol{\mu}}_{H}\left(t\right)\hat{\boldsymbol{\mu}}_{H}\left(0\right)\right\rangle \nonumber \\ 
 & =\text{\ensuremath{\left|\boldsymbol{\mu}\right|}}^{2}\left\langle e^{i\left(E_{1}+H_{b}+\Delta\right)t/\hbar}e^{i\left(E_{g}+H_{b}\right)t/\hbar}\right\rangle _{B} \nonumber\\ 
 & =\text{\ensuremath{\left|\boldsymbol{\mu}\right|}}^{2}e^{i\left(E_{1}-E_{g}\right)t/\hbar}\cdot\left\langle F_1\left(t\right)\right\rangle \,, \label{eq:trace}
\end{align}
where the dephasing function between excitonic state $\left|\psi_{1}\right\rangle$ and the ground state is $\left\langle F_1\left(t\right)\right\rangle \equiv\left\langle \exp_{T}\left\{ \frac{i}{\hbar}\int_{0}^{t}d\tau\Delta\left(\tau\right)\right\} \right\rangle$. And both $H_{b}=\sum_{\alpha}\hbar\omega_{\alpha}b_{\alpha}^{\dagger}b_{\alpha}$
and $\Delta=\sum_{\alpha}V_{1,1}^{\alpha}q_{\alpha}$ are operators
in the nuclear (phonon) Hilbert space. $\left\langle \cdot\right\rangle =Tr_{b}\left[\rho_{b}\cdot\right]$. 
$\Delta\left(\tau\right)=e^{i\left(H_{b}\right)t/\hbar}\Delta e^{-i\left(H_{b}\right)t/\hbar}$
is the Heisenberg representation of the coupling operator. The subscript $T$ keeps the operators in decreasing time order. 

A cumulant expansion on the dephasing function is then performed.\cite{Skinner1986,Egorov1997a,Egorov1996,Egorov1995} Since the operator $\Delta$ is linear in the phonon coordinates, only the second cumulant is non-zero, and is expressed as: 
\begin{equation}
K_{2}\left(t\right) =\left(-\frac{i}{\hbar}\right)^{2}\int_{0}^{t}d\tau_{2}\int_{0}^{t-\tau_{2}}d\tau_{1} \sum_{\alpha,\beta}V_{1,1}^{\alpha}V_{1,1}^{\beta}\left\langle q_{\alpha}\left(\tau_{1}\right)q_{\beta}\left(0\right)\right\rangle \,, 
\label{eq:second-cumulant}
\end{equation}
Assuming all the phonons are harmonic oscillators and treating them quantum mechanically, the bath correlation function can be evaluated analytically.
\begin{equation}
\left\langle q_{\alpha}\left(t\right)q_{\beta}\left(0\right)\right\rangle =\frac{\hbar}{2\omega_{\alpha}}\left[\left(n\left(\omega_{\alpha}\right)+1\right)e^{-i\omega_{\alpha}t}+n\left(\omega_{\alpha}\right)e^{i\omega_{\alpha}t}\right]\delta_{\alpha\beta} \,, 
\end{equation}
where $n\left(\omega_{\alpha}\right)=\frac{1}{\exp\left(\beta\hbar\omega_{\alpha}\right)-1}$ is the Bose-Einstein distribution. 

Thus, our model gives an analytical expression for the emission spectrum between the lowest exciton state to the ground state.\cite{Skinner1986,Egorov1997a,Egorov1996,Egorov1995}
\begin{equation}
I_{1\rightarrow g}\left(\omega\right) =\frac{\text{\ensuremath{\left|\boldsymbol{\mu}_{g1}\right|}}^{2}}{2\pi}\int_{-\infty}^{\infty}dte^{-i\left(\omega-\frac{E_{1}-E_{g}}{\hbar}\right)t} \exp\Bigg\{\frac{1}{2\hbar}\sum_{\alpha}\frac{\left(V_{1,1}^{\alpha}\right)^{2}}{\left(\omega_{\alpha}\right)^{3}}\left[ C_{\alpha}^{\Re} (t) + iC_\alpha^{\Im} (t)\right]\Bigg\} \,,
\end{equation}
where $C_{\alpha}^{\Re} (t)$ and $C_\alpha^{\Im} (t)$ are given in the main text. Our model is then easily extended to multiple electronic transitions, the effect of which only becomes noticeable when the temperature is high. In PL emission spectroscopy, the initial excitonic states are populated according to a Boltzmann thermal distribution, and the strength of the spectrum is proportional to the transition dipole moment. Therefore, the spectra of each exciton-ground-state transition are summed up with the corresponding weights, as expressed below: 
\begin{equation}
I\left(\omega\right) =\frac{1}{Q_{n}}\sum_{n}e^{-\beta E_{n}}\left|\boldsymbol{\mu}_{gn}\right|^{2}\int_{-\infty}^{\infty}dte^{i\left(\omega-\omega_{ng}\right)t} \exp\Bigg\{\frac{1}{2\hbar}\sum_{\alpha}\frac{\left(V_{n,n}^{\alpha}\right)^{2}}{\left(\omega_{\alpha}\right)^{3}}\left[ C_{\alpha}^{\Re} (t) + iC_\alpha^{\Im} (t)\right]\Bigg\} \,, 
\label{eq:final_analytical_spectrum}
\end{equation}
where a sum over all excitonic states $n$ is performed, $E_{n}$ is the energy of state $n$, and $\omega_{gn}=\frac{E_{n}-E_{g}}{\hbar}$. $\left|\boldsymbol{\mu}_{gn}\right|$ is the transition dipole moment between exciton state $n$ and the ground state, calculated by the following integral on the single-particle wavefunctions. $e$ is the electron charge. 
\begin{equation}
\left|\boldsymbol{\mu}_{gn}\right|^{2}=\left\langle \psi_{g}\left|e\mathbf{r}\right|\psi_{n}\right\rangle=e\sum_{ai} c_{a,i}^{n} \int d\mathbf{r} \phi_a\left( \mathbf{r}_e\right) \mathbf{r} \phi_i\left( \mathbf{r}_h\right) \,,
\label{eq:transition-dipole-element}
\end{equation}

Obtaining the FWHM from the analytical calculations of the spectra is tricky, since our model Hamiltonian lacks a dephasing time~\cite{Skinner1986} and exhibits recurrences due to the finite number of phonon modes, resulting in infinitely narrow vibronic peaks (see main text Fig.1(b)). In order to reveal the intrinsic dynamics of the electronic transition and calculate experimentally-relevant FWHM, we convolved the calculated vibronic spectra with a time-domain Gaussian filter function with a width of $\sim 4$meV before comparing with experimental single-molecule PL measurements. This width is also consistent with the instrument resolution limit of the experimental setup, and the timescale of the decay of this filter function is faster than the recurrence time, $t_r \approx 2\pi/\omega_{\rm min}$. Furthermore, to match ensemble measurements, we convolved the calculated spectrum with another (broader) Gaussian function to account for the nearly-normal distribution of inhomogeneity of local environments in ensemble PL experiments.

\section{Anharmonicity and PL linewidth}
We investigated whether anharmonicity of atomic motion at high temperatures, manifested through the finite lifetimes of phonons, affects the width of the PL spectrum. We concluded that the finite lifetimes of phonons only broadens the FWHM of PL spectra by around $10$meV at $300$K, and is thus insufficient to explain the mismatch in FWHM - temperature relationship between our linear exciton-phonon coupling model and the experimental single-NC measurements. We then empirically correct the dephasing rate by accounting for higher-order expansion terms in the exciton-phonon couplings, and analyze these additional decay channels (as discussed in the main text). Here, the details of anharmonicity are given. 

The finite lifetime of the phonons were calculated following the methods outlined in Ref.~\citenum{Guzelturk2021}. The relaxation lifetimes $1/T_1$ of each phonon mode in a $3.0$ nm core, $3$ML shell CdSe/CdS nanocrystal were calculated from the velocity auto-correlation function $C(t)=\langle v(0)v(t)\rangle$ using the following relation:\cite{Guzelturk2021}
\begin{equation}
\frac{1}{T_1} = \tilde{\xi}^\prime (\omega) =\Re \left[ \frac{C(0)-i\omega \tilde{C}(\omega)}{\tilde{C}(\omega)} + i\omega\right] \,, 
\end{equation}
where $\tilde{\xi}^\prime (\omega)$ is the real part of the Laplace transform of the friction kernel from the generalized Langevin equation at the frequency of the mode. The velocity auto-correlation functions were calculated directly by simulating molecular dynamics (MD) trajectories which were sampled from a microcanonical ensemble at $300$K. The same Stillinger-Weber force field parameterized for bulk CdSe and CdS (as described in the main text and earlier sections of the SI) was used. The transformation from atomic coordinates to phonon mode coordinates was done through the eigenvectors of the mass-weighted Hessian matrix. The lifetimes of phonons that couple most strongly to the ground bright excitonic state have the most significant effects on the dephasing dynamics. As shown in Fig.2e of Ref.~\citenum{Guzelturk2021}, the MD simulations reveal that the most strongly coupled acoustic phonons at $\sim 0.5$ THz have lifetimes of about $1$ ps, while the CdSe optical phonon modes at $\sim 8$ THz have lifetimes of around $100$ fs. 

The finite phonon lifetime is incorporated into the bath correlation function by considering its damping effect on each phonon motion. The equations of motion of the creation and annihilation operators in these quantum mechanical, damped harmonic oscillators can be rewritten as: 
\begin{equation}
\dot{\hat{b}}_\alpha = -i\omega_{\alpha}\hat{b}_\alpha - \frac{1}{T_1}\hat{b}_\alpha \,, \quad h.c. \,,
\end{equation}
where $\hat{b}_\alpha$ and $\hat{b}^\dagger_\alpha$ are the annihilation and creation operators for phonon mode $\alpha$ with frequency $\omega_{\alpha}$ and lifetime $T_1$. For simplicity, we assume that all phonons have the same lifetime and thus, the bath correlation function for the damped harmonic oscillator is exponentially damped. 
\begin{equation}
\langle q_{\alpha}\left(t\right)q_{\beta}\left(0\right)\rangle = e^{-\frac{t}{T_1}}\langle q_{\alpha}\left(t\right)q_{\beta}\left(0\right)\rangle_{HO} \,,
\end{equation}
From here, we can plug the damped bath correlation function into Eq.~\eqref{eq:second-cumulant} to obtain the analytical PL spectra: 
\begin{equation}
I\left(\omega\right) =\frac{1}{Q_{n}}\sum_{n}e^{-\beta E_{n}}\left|\mu_{gn}\right|^{2}\int_{-\infty}^{\infty}dte^{i\left(\omega-\omega_{ng}\right)t} \exp\Bigg\{\frac{1}{2\hbar}\sum_{\alpha}\frac{\left(V_{n,n}^{\alpha}\right)^{2}}{\left(\omega_{\alpha}\right)^{3}} C_{\alpha}^{DHO} (t)\Bigg\} \,,
\end{equation}
where for the each damped harmonic oscillators, 
\begin{align}
C_{\alpha}^{DHO} (t) = \omega_{\alpha}^2 \cdot \Bigg[ & \left( n( \omega_{\alpha})+1\right) \frac{e^{-t/T_1 - i\omega_{\alpha}t}+\frac{t}{T_1}+i\omega_{\alpha}t-1}{(\frac{1}{T_1}+i\omega_{\alpha})^2} \nonumber\\
&+ n( \omega_{\alpha})\frac{e^{-t/T_1 + i\omega_{\alpha}t}+\frac{t}{T_1}-i\omega_{\alpha}t-1}{(\frac{1}{T_1}-i\omega_{\alpha})^2}\Bigg] \,,
\end{align}

Using the phonon lifetime $T_1\approx 100$fs at $300$K for $3$nm CdSe / $3$ML CdS core-shell nanocrystal, we obtained the anharmonic correction to the PL spectrum. At $300$K, incorporating anharmonicity increases the model PL FWHM from $28.2$meV (for harmonic phonons) to $37.8$meV. This correction of $\approx 10$meV in the single-NC FWHM only accounts for a very small fraction of the discrepancy between the experimental linewidth ($68.10$meV) and our model with linear exciton-phonon coupling($28.2$meV). Thus, we ruled out anharmonicity as the primary mechanism for the temperature-activated emission channel, and investigated high-order exciton-phonon coupling expansions, as discussed in the main text. 

\bibliography{main}